\documentclass[11pt,a4paper]{article}
\date{}
\usepackage{amsmath,amsthm}
\usepackage{graphicx}

\graphicspath{ {./images/} }
\newtheorem*{t1}{Theorem 1}
\newtheorem*{t2}{Theorem 2}
\newtheorem*{t3}{Theorem 3}
\newtheorem*{t4}{Theorem 4}
\newtheorem*{t5}{Theorem 5}
\newtheorem*{c1}{Corollary 1}
\newtheorem*{l1}{Lemma 1}
\newtheorem*{l2}{Lemma 2}

\begin{document}

\markboth{Ch. Dangalchev}
{Closeness and Decision Making}

\title{Closeness and Decision Making}

\author{Chavdar Dangalchev \\ dangalchev@hotmail.com
   \\ Institute of Mathematics and Informatics
   \\  Bulgarian Academy of Sciences, Sofia, Bulgaria. }

\maketitle

\begin{abstract}

In this article we consider networks, which for a given
time period can have one link broken. Which new link
should we build so the  closeness of the resulting
network satisfies some optimal criteria?

We consider different criteria for optimization and
different graphs - cycle, paths, lollipop graphs,
and two complete graphs, connected by a link.


\end{abstract}

\section{Introduction}

There are two major components of the analysis of complex networks - growth and vulnerability. 
One of the most important centrality measures of graphs is closeness.
Dangalchev in a work for network vulnerability [1] 
proposed one of the most sensitive characteristics - residual closeness. 
It measures the closeness of a graph after deleting a vertex or a link. The  definition for
 graph closeness  in simple undirected graphs, used in [1], is:
\begin{equation}
\label{eq1}
C(G)=\sum\limits_i  \sum\limits_{j\ne i} 2^{-d(i,j)}. 
\end{equation}
\noindent  In the above formula  $d(i,j)$  is the standard distance between vertices $i$  
and $j$. 
The advantages of the above definition are that it can be used 
 for not connected graphs, and it is convenient for creating formulae for graph operations. 

Let $i$   and $j$  be a pair of connected vertices of graph  $G$ and graph  $G_{i,j}$  be the graph, 
constructed by removing link $(i,j)$ from  $G$. 
Let  $ d_{i,j} (r,s)$  be the distance between vertices
$r$   and $s$  in  graph  $G_{i,j}$.
Using  formula (1), we can  calculate the closeness of graph   $G_{i,j}$:
\[
C (G_{i,j})  =  \sum\limits_r {\sum\limits_{s\ne r} {2^{-d_{i,j}(r,s)}} }.
\]

The link residual closeness R, a measure of graph $G$ vulnerability, is defined in [1] as:
\[
R(G)={\min_{i,j} \{ C(G_{i,j}) \}  }.
\]

Let $p$   and $q$  be a pair of not connected vertices in graph  $G$ and  
graph  $G_{p,q}$  be the graph, constructed by connecting $p$ and $q$. 
The additional closeness is a measure of graph $G$ growth potential,
and it is defined in Dangalchev [2] as:
\[
A(G)={\max_{p,q} \{ C(G_{p,q}) \}  }.
\]

The additional and the residual closeness describe the behavior of graphs in two different directions - building and destroying the graph.

Let us consider a situation where we have limited resources for maintaining a network (e.g., railroad system, water distribution system, electrical system, etc.).  Where should we focus our efforts:
on expanding the network
by building new links or preventing the existing lines from failure? 
Everything depends on the network structure and the cost of building and preventing failures. A simple answer can give us a comparison between the residual and the additional closeness 
(e.g., by adding normalized, divided by the closeness,  additional and residual closenesses [3]).

\setlength{\unitlength}{.45in}
\begin{picture}(11,3.5)(-0.0,-1.5)
\linethickness{0.7pt}


\put(3,0){\circle{0.08}}
\put(3,1){\circle{0.08}}
\put(1,0){\circle{0.08}}
\put(2,0){\circle{0.08}}
\put(1,1){\circle{0.08}}

\linethickness{0.7pt}

\put(2,0){\line(-1,1){1.0}}
\put(2,0){\line(1,1){1.0}}

\linethickness{0.6pt}
\put(1,0){\line(0,1){1}}

\put(3,0){\line(0,1){1}}
\put(1,0){\line(1,0){2}}

\put(2,-1.0){\makebox(0,0){Fig. 1: Graph $G_1$}}


\put(3.5,0){\circle{0.08}}
\put(3.5,1){\circle{0.08}}
\put(4.5,0){\circle{0.08}}
\put(5.5,0){\circle{0.08}}
\put(6.5,0){\circle{0.08}}
\put(6.5,1){\circle{0.08}}

\put(4.5,0){\line(-1,1){1}}
\put(5.5,0){\line(1,1){1}}

\linethickness{0.6pt}
\put(3.5,0){\line(0,1){1}}

\put(6.5,0){\line(0,1){1}}
\put(3.5,0){\line(1,0){3}}

\put(4.5,-0.3){\makebox(0,0){$1$}}
\put(5.5,-0.3){\makebox(0,0){$2$}}

\put(5.1,-1.0){\makebox(0,0){Fig. 2: Graph $G_2$}}


\put(7,0){\circle{0.08}}
\put(7,1){\circle{0.08}}
\put(8,0){\circle{0.08}}
\put(9,0){\circle{0.08}}
\put(10,0){\circle{0.08}}
\put(10,1){\circle{0.08}}

\put(8,0){\line(-1,1){1}}
\put(9,0){\line(1,1){1}}

\linethickness{0.6pt}
\put(7,0){\line(0,1){1}}
\put(10,0){\line(0,1){1}}
\put(7,0){\line(1,0){1}}
\put(9,0){\line(1,0){1}}

\put(8,-0.3){\makebox(0,0){$1$}}
\put(9,-0.3){\makebox(0,0){$2$}}

\put(8.6,-1.0){\makebox(0,0){Fig. 3: Graph $G_3$}}

\end{picture}

Let us compare the graphs from Fig. 1, Fig. 2, and Fig. 3.
The closeness (C), residual closeness(R), and additional closeness (A)
of graphs $G_1$, $G_2$, and $G_3$ are: 
\[
C(G_1) = 8.0, \quad \quad R(G_1) = 7.0 , \quad \quad A(G_1) = 8.5,
\]
\[
C(G_2) = 10.0, \quad \quad R(G_2) = 6.0 , \quad \quad A(G_2) = 11.25,
\]
\[
C(G_2) = 6.0, \quad \quad R(G_2) = 5.5 , \quad \quad A(G_2) = 10.0.
\]

We can calculate the normalized residual closeness: $NR(G) = R(G) / C(G)$ and normalized additional closeness: $NA(G) = A(G) / C(G)$.
By calculating the normalized residual closeness, we will have an indicator
of graph vulnerability. We can see that
the graphs $G_1$ and $G_3$ are very robust ($NR(G_1) = 87.5 \%$ and 
$NR(G_3) = 91.67 \%$), 
while graph $G_2$ is too vulnerable ($NR(G_2) = 60.0\%$).
We should put more efforts protecting  link $(1,2)$ of graph $G_2$.

The normalized additional closeness shows the growth potential of networks.
Graphs $G_1$ and $G_2$ have little growth potential  ($NA(G_1) = 106.25 \%$ and $NR(G_2) = 112.5 \%$), 
while graph $G_3$ has big potentials ($NR(G_3) = 166.67\%$).
If we have the option to build one link in graph $G_3$, this should be a link, connecting
the two parts in Fig. 3 (e.g., link $(1,2)$).

Some other results about closeness, residual closeness, and
additional closeness can be found in ([6] - [24]).

We consider a situation where one link of a graph is randomly broken and we could build 
(add) another link. We use decision making theory to find the best link to add to increase 
the closeness.  In this paper we do this for some specific graph classes: cycles, lollipop graphs,
 complete graphs connected by a single edge.

\section{Decision Theory - Criteria}

Let us maintain a network with options to protect links or build new ones. 
It looks as if we play a (non-cooperative, non-zero-sum and simultaneous) game against nature - some existing links (lines, pipes) can break, and we can build new links. If for a given period, we build one link and one link breaks then the closeness of the resulting network looks like a payoff table (see Table 1 below). One way to treat this problem is to apply game theory (see for example [4]) or decision theory (see [5]) .

A simple example of a payoff table gives us path graph $P_{4}$ with vertices \{1, 2, 3, 4\} and closeness $C(P_{4}) = 4.25$. 
Below is a table with the closenesses, received after deleting one existing link 
($(1,2)$, $(2,3)$, or $(3,4)$) 
and adding one possible link ($(1,3)$, $(1,4)$, or $(2,4)$) :

 \quad 

\begin{table}[h!]
  \begin{center}
    \begin{tabular}{|c|c c c|} 
      \hline
      \textbf{Del - Add} & \textbf{(1,3)} & \textbf{(1,4)} & \textbf{(2,4)}\\
      \hline
 \textbf{(1,2)}   & 4.5	 &  4.25 & 3\\
\textbf{(2,3)}    & 4.25 &  4.25 & 4.25\\
\textbf{(3,4)}    & 3	 & 4.25 & 4.5\\
      \hline
    \end{tabular}
  \end{center}
\caption{\label{tab:table-name}Payoff table of path $P_4$.}
 \end{table}

The point in the middle of Table 1 ($4.25$) is a saddle point and 
 Nash equilibrium point  - we should build link $(1,4)$ to ensure closeness  $4.25$. Most of the payoff tables do not have saddle points. We need to find another way to determine our best action.

Let us have a graph $G$.  Let for a given period one link $(i,j)$ of $G$ break and we can build another link $(p,q)$. There are different criteria to evaluate 
 the best strategy to pick  link $(p,q)$. In all the criteria below, we are trying to maximize  the closeness of graph $G'$, received after deleting link $(i,j)$ and adding link $(p,q)$.
 After deleting link $(i,j)$ from graph $G$ we receive graph $G_{i,j}$ with closeness $C(G_{i,j})$. Adding to graph $G_{i,j}$ link $(p,q)$ we receive graph $G^{p,q}_{i,j}$. We will maximize the payment function  $P(p,q)$, which is a function of the closeness of graph $G^{p,q}_{i,j}$. 
There are different criteria, different function  $P(p,q)$, to be maximized:

1. \textbf{Equal Likelihood}:

If all nature outcomes (deleting a link $(i,j)$) are with equal likelihood, we can calculate:
\[
P(p,q) = \sum\limits_{i,j} C(G^{p,q}_{i,j}).
\]
Using it, we can calculate the average closeness $Av$.

2. \textbf{Different probability}:

If there are different probabilities $p(i,j)$ for every deleted link $(i,j)$, then:
\[
P(p,q)  = \sum\limits_{i,j} p(i,j) \cdot C(G^{p,q}_{i,j}).
\]

Most of the time the probabilities are not known - we need to make a choice under uncertainty. We should use the following strategies.

3. \textbf{Pessimistic  strategy} (trying to avoid big loss):
\[
P(p,q)={\min_{i,j} \{  C(G^{p,q}_{i,j}) \}}.
\]
In this case we can denote the maximum of the minimum of the payment function by $Mn(G)$:
\[
Mn(G)={\max_{p,q} \{  P(p,q)  \} }= {\max_{p,q} \{ {\min_{i,j} \{  C(G^{p,q}_{i,j}) \}} \} }.
\]

4. \textbf{Optimistic strategy} (targeting the big gain):
\[
P(p,q)={\max_{i,j} \{  C(G^{p,q}_{i,j}) \}}.
\]
We can denote the maximum of the maximum by $Mx(G)$:
\[
Mx(G)={\max_{p,q} \{ {\max_{i,j} \{  C(G^{p,q}_{i,j}) \}} \} }.
\]

5. \textbf{Hurwitz criterion}:

It is a mix of the above two strategies - we
give different weights on pessimistic and optimistic criteria:
\[
P(p,q)= \alpha \cdot   {\max_{i,j} \{  C(G^{p,q}_{i,j}) \}} + (1-\alpha) {\min_{i,j} \{  C(G^{p,q}_{i,j}) \}},
\]
where $ 0 < \alpha <1$. Extreme values give us pessimistic and optimistic criteria.
In case of $\alpha = 0.5$ we can use for calculations two times the payment function or
$Mn(G)+Mx(G)$.

6. \textbf{Savage criterion}:

We create a table with regrets (how much the payoff $C(G^{p,q}_{i,j})$ is different than the maximal possible 
$\max_{i,j} \{  C(G^{p,q}_{i,j}) \}$ and we try to minimize the regrets.
\[
Reg = P(p,q)=-{\max_{p,q} \{ {\max_{i,j} \{  C(G^{p,q}_{i,j}) \}}   - C(G^{p,q}_{i,j})   \}}.
\]

\section{Additional closeness vs. Decision Making}

Why do we need to use decision making techniques?
Can we use (build) the link, giving the additional closeness? 
The  additional closeness is the maximal closeness we can add to the graph. The answer is that the link supplying the additional closeness is not always (by all the above criteria) the best decision.

Let us consider cycle graph $C_6$ (see [3]) with  vertices $\{1,2,3,4,5,6\}$. The closeness $C(C_6) = 9.75$. The residual closeness is $R(C_6) = 8.0625$, received by deleting any link. The additional closeness is $A(C_6) = 10.75$, received by adding for example link $(1,3)$. By adding links connecting opposite vertices, for example link $(1,4)$ we receive closeness $10.5$.

Adding links connecting opposite vertices (e.g., $(1,4)$)  gives us maximal closeness (optimistic criteria) $Mx= 9.75$, minimal closeness
(pessimistic criteria) $Mn = 9.375$, Hurwitz criteria with coefficients 0.5 is 
$Mx + Mn = 19.125$,
average of all closnesses $avg(C(C_6')) = 9.5$, and Savage criteria - maximal regret $reg = 0.375$.

Adding not opposite vertices (e.g. $(1,3)$)  give us maximal closeness
$Mx= 10$, minimal closeness $Mn= 9$, $Mx + Mn = 19$, average of all closnesses $avg(C) = 9.45833$, and maximal regret $reg = 1$.

We can see that linking not opposite vertices  (e.g. link $(1,3)$, which supplies the additional closeness) is the best only by optimistic strategy. 
Adding opposite links  (e.g. link $(1,4)$) is best by pessimistic strategy, equal likelihood, Hurwitz criteria (for $\alpha = 0.5$) , and minimal regret. This is the reason for considering decision making techniques.

\section{Previous Results}

The closeness of path graph $P_n$ with $n$ vertices , proven in [1] is:
\begin{equation}
\label{eq2}
C(P_n) = 2n -4  + 2^{2-n} .
\end{equation}

The closeness of cycle graph $C_n$ with $n$ vertices , proven in [1] and [6] is:
\begin{equation}
\label{eq3}
C(C_n) = \left \{
  \begin{array}{lr}
     4k(1-3.2^{-k-1} ),   \quad when \quad  n = 2k,\\
     2(2k+1)(1-2^{-k} ) ,     \quad when \quad n = 2k+1.
  \end{array}
\right .
\end{equation}

One advantage of the definition (1) is that it is convenient for creating formulae for graph operations.  
For example, if we connect vertex $p$ of graph $G_1$ with vertex 
$q$ of graph $G_2$ the closeness of the connected graph $G_1 + G_2$ is ([1]):

\begin{equation}
\label{eq4}
C(G_1 + G_2)=C(G_{1} )+C(G_{2} )+\left( {1+C(p)} \right)\left( {1+C(q)} \right).
\end{equation}

\section{Decision making for two complete graphs}

Let two complete graphs ($K_k$ and $K_m$)
with $k$ and $m$ vertices ($k,m \ge 3$) be connected with a link to create graph $G=K_k+K_m$. This is a generalization of the example of Table 1: $P_4=K_2+K_2$.
We will find the best link to be built by all criteria from section 3.
At the beginning we will prove:

\begin{l1}
The closeness of graph $K_k+K_m$ is:
\[
C(G)=C(K_k + K_m) =  \frac{2k^2+2m^2 +km -k -m+1}{4}.
\]
\end{l1}

\begin{proof} 
Using formula (4) we can calculate:

\begin{align}
\begin{split}\label{}
C(K_k + K_m)  & = \frac{k(k-1)}{2} +  \frac{m(m-1)}{2} + 
\left (1+ \frac{k-1}{2} \right )   \left (1+  \frac{m-1}{2}  \right )
\nonumber 
 \\&  = \frac{k^2-k +m^2 -m}{2}+ \frac{k m+k +m+1}{4}
 \\&  = \frac{2k^2+2m^2 +km -k -m+1}{4}.
\end{split}
\end{align}
which finishes the proof.
\end{proof} 

Let the vertices of graph $K_k+K_m$ are ${1,2,...,k}$ of complete graph $K_k$ and
${k+1,k+2,...,k+m}$ of complete graph $K_m$. Let the link between $K_k$ and $K_m$
be $(1,k+m)$.

\setlength{\unitlength}{.45in}
\begin{picture}(11,3.5)(-0.0,-1.5)


\put(0.5,0){\circle{0.08}}
\put(0.5,1){\circle{0.08}}
\put(1.5,0){\circle{0.08}}
\put(1.5,1){\circle{0.08}}
\put(2.5,0){\circle{0.08}}
\put(2.5,1){\circle{0.08}}
\put(3.5,1){\circle{0.08}}

\linethickness{0.7pt}
\put(0.5,0){\line(1,1){1}}
\put(0.5,1){\line(1,-1){1}}
\put(2.5,0){\line(1,1){1}}
\put(1.5,0){\line(0,1){1}}
\put(2.5,0){\line(0,1){1}}

\linethickness{0.6pt}
\put(0.5,0){\line(0,1){1}}
\put(2.5,1){\line(1,0){1}}

\put(0.5,0){\line(1,0){2}}
\put(0.5,1){\line(1,0){3}}

\put(0.5,-0.3){\makebox(0,0){$2$}}
\put(1.5,-0.3){\makebox(0,0){$1$}}
\put(2.5,-0.3){\makebox(0,0){$k+m$}}
\put(0.5, 1.3){\makebox(0,0){$3$}}
\put(1.5, 1.3){\makebox(0,0){$k$}}
\put(2.5, 1.3){\makebox(0,0){$k+1$}}

\put(2.1,-1.0){\makebox(0,0){Fig. 4: Graph $G_4$}}


\put(6.5,0){\circle{0.08}}
\put(6.5,1){\circle{0.08}}
\put(7.5,0){\circle{0.08}}
\put(7.5,1){\circle{0.08}}
\put(8.5,0){\circle{0.08}}
\put(8.5,1){\circle{0.08}}
\put(9.5,1){\circle{0.08}}

\linethickness{0.7pt}
\put(6.5,0){\line(1,1){1}}
\put(6.5,1){\line(1,-1){1}}
\put(7.5,0){\line(1,1){1}}
\put(8.5,0){\line(1,1){1}}
\put(7.5,0){\line(0,1){1}}
\put(8.5,0){\line(0,1){1}}

\linethickness{0.6pt}
\put(6.5,0){\line(0,1){1}}
\put(8.5,1){\line(1,0){1}}

\put(6.5,0){\line(1,0){2}}
\put(6.5,1){\line(1,0){1}}

\put(6.5,-0.3){\makebox(0,0){$2$}}
\put(7.5,-0.3){\makebox(0,0){$1$}}
\put(8.5,-0.3){\makebox(0,0){$k+m$}}
\put(6.5, 1.3){\makebox(0,0){$3$}}
\put(7.5, 1.3){\makebox(0,0){$k$}}
\put(8.5, 1.3){\makebox(0,0){$k+1$}}

\put(8.1,-1.0){\makebox(0,0){Fig. 5: Graph $G_5$}}

\end{picture}

In Figures 4 and 5 we can see two complete graphs $K_4$ and $K_3$,
connected with a link $(1,k+m)$ (in this case link $(1,7)$).
Any new link will connect  $K_k$ and $K_m$ again. 
There are two cases:
both newly connected vertices are different than vertices $1$ and $k+m$ (see Fig.4, graph $G_4$ - the added link is $(k,k+1)$) or one of the vertices
is the same (see Fig.5, graph $G_5$ - the added link is
$(1,k+1)$).

The link deleted could be the existing link between the two complete graphs, 
a link including  one of the vertices $1$ or $k+m$, or a different link. There are 5 cases:

 \quad  \textbf{CASE A:}
Let us add a new link $(k,k+1)$. If the deleted link is $(1,k+m)$ the newly
created graph $G'$ is the same as graph  $G$: $C(G')=C(G)$. The same is the situation if we
add link $(1,k+1)$ and the deleted link is $(1,k+m)$.

 \quad  \textbf{CASE B:} Let us add link $(k,k+1)$ (see Fig.4, graph $G_4$). Let the deleted link do not include vertices $1$, $k$, $k+1$, and $k+m$ (e.g., link 
$(2,3)$). There are $(k-2)(k-3) /2+(m-2)(m-3) / 2$ links like this. The closeness of graph $G'$ is different from the closeness of  graph  $G$. We will calculate this difference:

	- The distance between vertices $2$ and $3$ is changed from $1$ to $2$. In the graph 
$G'$ this closeness is calculated 2 times: for distances $d(2,3)$ and $d(3,2)$.

	- The distance between vertices $k$ and $k+1$ is changed from $3$ to $1$.

	- The distance between vertex $k$ and $(m-2)$ vertices from graph $K_m$ (excluding vertices $k+1$ and $k+m$)
is changed from $3$ to $2$.

	- The distance between vertex $k+1$ and $(k-2)$ vertices from graph $K_k$ (excluding vertices $1$ and $k$)
is changed from $3$ to $2$.

No other distances are changed.
The closeness of graph $G'$ (with added link $(k,k+1)$ and deleted link $(2,3)$) is:

\begin{align}
\begin{split}\label{}
C(G')  & = C(G)  + 2 \left ( \frac{1}{4} - \frac{1}{2}\right)+ 2 \left ( \frac{1}{2} - \frac{1}{8}\right)
\nonumber 
 \\&   \quad  + 2 (m-2) \left ( \frac{1}{4} - \frac{1}{8}\right)
+ 2 (k-2) \left ( \frac{1}{4} - \frac{1}{8}\right)
 \\&  = C(G) - \frac{1}{2} + \frac{3}{4}  + \frac{m-2}{4} + \frac{k-2}{4} 
 \\&  = C(G)+ \frac{k+m-3}{4}.
\end{split}
\end{align}
If $k = 3$ or $m = 3$, case B is impossible, and it is considered below.

 \quad  \textbf{CASE C:}  Let us add link $(k,k+1)$ (see Fig.4).  Let the deleted link include one vertex $1$, $k$, $k+1$, or $k+m$ (e.g., link $(1,2)$). There are $2k+2m-8$ links like this. We can calculate the difference between graphs $G'$ and $G$:

	- The distance between vertices $1$ and $2$ is changed from $1$ to $2$. 

	- The distance between vertices $k$ and $k+1$ is changed from $3$ to $1$.

	- The distance between vertices $2$ and $k+m$ is changed from $2$ to $3$.

	- The distance between vertex $k$ and $(m-2)$ vertices from graph $K_m$ (excluding vertices $k+1$ and $k+m$)
is changed from $3$ to $2$.

	- The distance between vertex $k+1$ and $(k-2)$ vertices from graph $K_k$ (excluding vertices $1$ and $k$) is changed from $3$ to $2$.

No other distances are changed.
The closeness of graph $G'$ (with added link $(k,k+1)$ and deleted link $(1,2)$) is:
\begin{align}
\begin{split}\label{}
C(G')  & = C(G)  + 2 \left ( \frac{1}{4} - \frac{1}{2}\right)+ 2 \left ( \frac{1}{2} - \frac{1}{8}\right)
+ 2 \left ( \frac{1}{8} - \frac{1}{4}\right)
\nonumber 
 \\&   \quad  + 2 (m-2) \left ( \frac{1}{4} - \frac{1}{8}\right)
+ 2 (k-2) \left ( \frac{1}{4} - \frac{1}{8}\right)
\end{split}
\end{align}

\begin{align}
\begin{split}\label{}
C(G')  &  = C(G) - \frac{1}{2} + \frac{3}{4} - \frac{1}{4}  + \frac{m-2}{4} + \frac{k-2}{4} 
\nonumber 
 \\&  = C(G)+ \frac{k+m-4}{4}.
\end{split}
\end{align}

 \quad  \textbf{CASE D:}  Let us add link $(k,k+1)$ (see Fig.4).  Let the deleted link be $(1, k)$. There are $2$ links like this ($(1, k)$ and 
$(k+1,k+m)$). We can calculate this difference between graphs $G'$ and $G$:

	- The distance between vertices $1$ and $k$ is changed from $1$ to $2$. 

	- The distance between vertices $k$ and $k+1$ is changed from $3$ to $1$.

	- The distance between vertices $k$ and $k+m$ is not changed 
(the distance is $2$).

	- The distance between vertex $k$ and $(m-2)$ vertices from graph $K_m$ (excluding vertices $k+1$ and $k+m$)
is changed from $3$ to $2$.

	- The distance between vertex $k+1$ and $(k-2)$ vertices from graph $K_k$ (excluding vertices $1$ and $k$)
is changed from $3$ to $2$.

No other distances are changed.
The closeness of graph $G'$ (with added link $(k,k+1)$ and deleted link $(1,k)$) is:
\begin{align}
\begin{split}\label{}
C(G')  & = C(G)  + 2 \left ( \frac{1}{4} - \frac{1}{2}\right)+ 2 \left ( \frac{1}{2} - \frac{1}{8}\right)
\nonumber 
 \\&   \quad  + 2 (m-2) \left ( \frac{1}{4} - \frac{1}{8}\right)
+ 2 (k-2) \left ( \frac{1}{4} - \frac{1}{8}\right)
 \\&  = C(G) - \frac{1}{2} + \frac{3}{4}   + \frac{m-2}{4} + \frac{k-2}{4} 
 \\&  = C(G)+ \frac{k+m-3}{4}.
\end{split}
\end{align}

 \quad  \textbf{CASE ABCD:} To summarize the calculations when we add link $(k,k+1)$ (cases A, B, C, and D): 
the minimal closeness is $Mn = C(G)$, 
the maximal closeness is $Mx = C(G) + (k+m-3)*0.25$,
the minimal + maximal closenesses is $Mn + Mx= 2C(G) + (k+m-3)*0.25$, 
the maximal regret is $Reg = (k+m-3)*0.25$,
and the average closeness is:

\begin{align}
\begin{split}\label{}
Av  & =   (  ( (k-2)(k-3)+(m-2)(m-3) + 4) \frac{k+m-3}{8} 
\nonumber 
 \\&  \quad   +( 2k+2m-8) \frac{k+m-4}{4} )
 \\&  \quad / \left ( k(k-1) /2+ m(m-1) /2 +1 \right ) + C(G) 
 \\&  = C(G)+ \frac{(k^2-5k+m^2-5m+16)(k+m-3) }{4 ( k^2 -k + m^2 - m + 2)}
 \\&   \quad + \frac{( k+m-4) (k+m-4)}{ k^2 -k + m^2 - m + 2}
\\& =  C(G)+ \frac{(k^2-k+m^2-m+2)(k+m-3) }{4 ( k^2 -k + m^2 - m + 2)} 
 \\&   \quad + \frac{(-4k-4m+14)(k+m-3) }{4 ( k^2 -k + m^2 - m + 2)} + \frac{( k+m-4) (k+m-4)} {k^2 -k + m^2 - m + 2}
\\&  = C(G)+ \frac{(k+m-3)}{4}  + \frac{k^2+m^2 +2km-8k-8m+16}{k^2 -k + m^2 - m + 2}
 \\&   \quad + \frac{ -2k^2-2m^2 -4km+13k+13m-21}{2(k^2 -k + m^2 - m + 2)}
\\&  =  C(G)+ \frac{(k+m-3)}{4}  + \frac{11-3k-3m}{2( k^2 -k + m^2 - m + 2)}.
\end{split}
\end{align}

 \quad

 \quad  \textbf{CASE E:}  Let us add link $(1,k+1)$ (see Fig.5, graph $G_5$) and the deleted link do not include vertex $1$ (e.g., link $(2,3)$). The situation is the same for $(k-1)(k-2)/2$
links of graph $K_k$ and all $m(m-1)/2$
links of graph $K_m$
The difference is:

	- The distance between vertices $2$ and $3$ is changed from $1$ to $2$.

	- The distance between vertices $1$ and $k+1$ is changed from $2$ to $1$.

	- The distance between vertex $k+1$ and $(k-1)$ vertices from graph $K_k$ (excluding vertex $1$)
is changed from $3$ to $2$.

No other distances are changed.
The closeness of graph $G'$ (with added link $(1,k+1)$ and deleted link $(2,3)$) is:
\begin{align}
\begin{split}\label{}
C(G')  & = C(G)  + 2 \left ( \frac{1}{4} - \frac{1}{2}\right)+ 2 \left ( \frac{1}{2} - \frac{1}{4}\right) + 2 (k-1) \left ( \frac{1}{4} - \frac{1}{8}\right)
\nonumber 
 \\&  = C(G)+ \frac{k-1}{4}.
\end{split}
\end{align}

 \quad  \textbf{CASE F:} Let us add link $(1,k+1)$ (see Fig.5) and the deleted link is $(1,2)$. There are $k-1$ links
like this. The difference in closeness is:

	- The distance between vertices $1$ and $2$ is changed from $1$ to $2$. 

	- The distance between vertices $1$ and $k+1$ is changed from $2$ to $1$.

	- The distance between vertices $2$ and $k+m$ is changed from $2$ to $3$.

	- The distance between vertex $2$ and $(m-2)$ vertices from graph $K_m$ (excluding vertices $k+1$ and $k+m$)
is changed from $3$ to $4$.

	- The distance between vertex $k+1$ and $(k-2)$ vertices from graph $K_k$ (excluding vertices $1$ and $2$)
is changed from $3$ to $2$. The distance between vertex $k+1$ and $2$ is the same - equal to 3.

No other distances are changed.
The closeness of graph $G'$ (with added link $(1,k+1)$ and deleted link $(1,2)$) is:

\begin{align}
\begin{split}\label{}
C(G')  & = C(G)  + 2 \left ( \frac{1}{2} - \frac{1}{4}\right) + 2 \left ( \frac{1}{4} - \frac{1}{2}\right) + 2 \left ( \frac{1}{8} - \frac{1}{4}\right)
\nonumber 
 \\&   \quad  + 2 (m-2) \left ( \frac{1}{16} - \frac{1}{8}\right)
+ 2 (k-2) \left ( \frac{1}{4} - \frac{1}{8}\right)
 \\&  = C(G)  - \frac{1}{4}  - \frac{m-2}{8} + \frac{k-2}{4} 
 \\&  = C(G)+ \frac{2k - m-4}{8}.
\end{split}
\end{align}

Comparing the cases E and F:
\[
 C(G) + \frac{2k - m-4}{8} =  C(G) +   \frac{k}{4} - \frac{m+4}{8}
< C(G) +   \frac{k-1}{4}. 
\]

 \quad  \textbf{CASE AEF:} To summarize  the calculations (cases A, E, and F), when we add link $(1,k+1)$  the maximal closeness is:
\[
Mx = C(G) + 0.25 \cdot {\max \{  k - 1, m - 1 \}}, 
\]
the minimal closeness is:
\[
Mn = C(G) + 0.125 \cdot {\min \{  0, (2k-m-4), (2m-k-4) \}} , 
\]
the maximal regret ($Reg = Mx-Mn$), when $k \ge m$ is:
\begin{align}
\begin{split}\label{}
Reg  & =   0.25 (k - 1) - 0.125 \cdot {\min \{  0, 2m-k-4 \}}
\nonumber 
 \\&  = 0.125 (2k - 2) + 0.125 \cdot {\max \{  0, k+4- 2m\}}
 \\&  =  0.125 \cdot {\max \{ 2k - 2, 3k+2- 2m\}}
\end{split}
\end{align}

When $k \ge m$ and $k+4 > 2m$:
\[
Reg = 0.125 (3k+2- 2m), 
\]
when $k \ge m$ and $2m \ge k+4$:
\[
Reg = 0.25 (k - 1). 
\]
The average closeness is:
\begin{align}
\begin{split}\label{}
Av  & =   \left (  ( (k-1)(k-2) / 2+m(m-1) / 2 ) \frac{k-1}{4} +
(k-1) \frac{2k-m-4}{8} \right) 
\nonumber 
 \\&  \quad  / \left ( k(k-1) /2+ m(m-1) /2 +1 \right ) + C(G) 
 \\&  = C(G) + \frac{ (k^2-3k+2+m^2-m+2k-m-4) (k-1) }
{4(k^2 -k + m^2 - m + 2)} 
 \\&  = C(G) +  \frac{k-1}{4} 
-  \frac{(m+4)(k-1)}{4(k^2 -k + m^2 - m + 2)}.
\end{split}
\end{align}

Summarizing  all the above we can prove:
\begin{t1}
 The maximum of maximum for two linked complete graphs ($K_k + K_m$) is:
\[
Mx(K_k + K_m) = C(K_k + K_m) + (k+m-3)*0.25.
\]
The maximum is received by connecting two vertices from graphs $K_k$ and $K_m$,
which do not include vertices, already connecting $K_k$ and $K_m$.
\end{t1}
\begin{proof} The maximum of maximum of cases A,B,C,D and A,E,F is:
\begin{align}
\begin{split}\label{}
Mx & =   C(G) + 0.25 \cdot {\max \{  (k+m-3), 
{\max \{  k - 1, m - 1 \}}\}}
\nonumber 
 \\&  =   C(G) + 0.25 \cdot {\max \{  (k+m-3), k - 1, m - 1\}}
 \\&  =   C(G) + 0.25(k+m-3),
\end{split}
\end{align}
 and this completes the proof.
\end{proof} 

\begin{t2}
The maximum of minimum for two  linked complete graphs ($K_k + K_m$) is:
\[
Mn(K_k + K_m) = C(K_k + K_m).
\]
The maximum of minimum is received by connecting two vertices from graphs $K_k$ and $K_m$, which do not include  vertices, already connecting $K_k$ and $K_m$.
\end{t2}
\begin{proof} The minimal closeness in cases A,B,C,D  is $Mn=C(G)$,
the minimal closeness in cases A,E,F  is:
\[
Mn = C(G) + 0.125 \cdot {\min \{  0, (2k-m-4), (2m-k-4) \}} \le C(G), 
\]
 and this finishes the proof.
\end{proof} 

Using the results of theorems 1 and 2 we receive:
\begin{c1}
 The sum of maximum and minimum for two linked complete graphs ($K_k + K_m$) is:
\[
Mx(K_k + K_m) + Mn(K_k + K_m)= 2C(K_k + K_m) + (k+m-3)*0.25.
\]
\end{c1}
This is received by connecting two vertices from graphs $K_k$ and $K_m$,
which do not include  vertices, already connecting $K_k$ and $K_m$.

\begin{t3}
 The minimum of maximal regret for two linked complete graphs ($K_k + K_m$) is:
\[
Reg(K_k + K_m) = \left \{
  \begin{array}{lr}
      \frac {k-1}{4},   \quad when \quad  2k \ge 2m \ge k+4,\\
     \frac {3k+2- 2m}{8}, \quad when \quad  k \ge m \quad and \quad 
k +8 \le 4 m \le 2k+8,\\
    \frac {k +m-3}{4},   \quad when \quad  k \ge m \quad and \quad 
k +8 > 4 m.
  \end{array}
\right .
\]
In the last situation, the minimum is received by connecting two vertices from graphs $K_k$ and $K_m$,
which do not include vertices, already connecting $K_k$ and $K_m$.
\end{t3}
\begin{proof} We can compare the cases A,B,C,D and A,E,F.
When $k \ge m$ and $2m \ge k+4$:
\[
\frac {k - 1}{4} < \frac {k +m-3}{4}
\]
and the minimum of maximal regret is $Reg = 0.25(k-1)$.
This situation is true when $k=5$ and $m=5$; $k=7$ and $m=6$, etc.

When $k \ge m$ and $k+4 \ge 2m$:
\[
\frac {3k+2- 2m}{8} < \frac {k +m-3}{4} 
\]
\[
\frac {k+8- 4m}{8} < 0.
\]
We have $Reg = 0.125(3k+2- 2m)$, when it is true: 
\[
2k+8 \ge 4m > k+8
\] 
or $k=3$ and $m=3$; $k=5$ and $m=4$; $k=7$ and $m=4$
; $k=6$ and $m=5$.

In all other cases ( $k \ge m$  and  
$k +8 > 4 m$) we have: $Reg = 0.25(k+m- 3)$.
This is true when $k=5$ and $m=3$; $k=9$ and $m=4$, etc., 
when connecting two vertices from graphs $K_k$ and $K_m$,
which do not include vertices, already connecting $K_k$ and $K_m$. This completes the proof.
\end{proof}

\begin{t4}
The maximum of average  for two  linked complete graphs ($K_k + K_m$) is:
\[
Av(K_k + K_m) = C(G)+ \frac{(k+m-3)}{4}  + \frac{11-3k-3m}{2( k^2 -k + m^2 - m + 2)}.
\]
The average is received by connecting two vertices from graphs $K_k$ and $K_m$,
which do not include vertices, already connecting $K_k$ and $K_m$.
\end{t4}

\begin{proof} We can compare the  average closeness in cases A,B,C,D and A,E,F
by calculating  the difference $\Delta$:

\begin{align}
\begin{split}\label{}
\Delta & = C(G)+ \frac{(k+m-3)}{4}  + \frac{11-3k-3m}{2( k^2 -k + m^2 - m + 2)}
\nonumber 
 \\&  \quad  -C(G) - \frac{k-1}{4} 
+ \frac{(m+4)(k-1)}{4(k^2 -k + m^2 - m + 2)}.
 \\&  = \frac{(m-2)}{4} + \frac{22-6k-6m +km+4k-m-4}{4(k^2 -k + m^2 - m + 2)}
 \\&  = \frac{(m-2)}{4} + \frac{18-2k-7m +km}{4(k^2 -k + m^2 - m + 2)}.
 \\&  = \frac{(m-2)}{4} + \frac{(k-7)(m-2)+4}{4(k^2 -k + m^2 - m + 2)}.
\end{split}
\end{align}

If $k \ge 7$ then $\Delta >0$. In all other cases we can prove it by consider:
\[
\Delta_4 = 4 \Delta (k^2 -k + m^2 - m + 2)
\]
\begin{align}
\begin{split}\label{}
\Delta_4 & = (m-2)(k^2 -k + m^2 - m + 2) +(k-7)(m-2)+4
\nonumber 
 \\&  = (m-2)(k^2 -5) + (m-2)(m^2 - m )+4 > 0.
\end{split}
\end{align}
 and this finishes the proof.
\end{proof} 

To summarize all, we have proven that the maximum of maximum,
the maximum of minimum, the Hurwitz criterion, the maximum of averages 
(Equal Likelihood criterion), and 1 out of 3 cases (see Theorem 3)
for the minimum of maximal regret
are received by connecting two vertices from graphs $K_k$ and $K_m$,
which do not include vertices, already connecting $K_k$ and $K_m$.
In 2 out of 3 cases the minimum of maximal regret
is received by connecting the connected vertex of graph $K_k$ to 
to a not-connected vertex of graph $K_m$, when $k \ge m$.

\section{Decision Making for Lollipop Graph}

Another easy calculation is for the maximum of maximum of lollipop graphs.
Let lollipop graph $L_{n,m}$ be created by connecting complete graph $K_n$ with path $P_m$.
For the optimistic strategy, the maximum is when deleting a link from the complete graph $K_n$,
not connected to the  path $P_m$. Then the closeness is decreasing by $1 / 2$:
\[
P(p,q)={\max_{i,j} \{  C(L_{i,j}) \}} = C(L_{p,q}) -\frac {1}{2} .
\]
\[
{\max_{p,q} P(p,q)} ={\max_{p,q}  C(L_{p,q}) } -\frac {1}{2} = A(L)   -\frac {1}{2}.
\]
The optimal strategy is to add  link $(p,q)$, which is giving the additional closeness $A(L_{n,m})$ of  lollipop graph $L_{n,m}$ (see [21]).
The other criteria for lollipop graphs are not that easy to calculate.

The same way we can directly prove Theorem 1 from the previous section. Following case B we can prove for the additional closeness:
\[
A(K_k+K_m) = C(K_k+K_m) + \frac {k+m-3}{4} + \frac {1}{2} = C(K_k+K_m) + \frac {k+m-1}{4}.
\]

\section{Formula for closeness of cycles}

Not for all graphs is it so easy  to determine the optimal link, as in the examples above.
We will give an example by calculating the maximin of cycle graphs.
First we need some preliminary calculations. 

Let us connect 2 vertices of cycle $C_m$ with  $m$ vertices $V = \{1,2,...,m \}$ and delete one link.
We receive graph $G$: cycle $C_{n}$ and two tails (paths) with $p$ and $q$ vertices, each connected to two neighboring vertices of the cycle $C_{n}$  ($m = n + p +q$).

\begin{l2}
The closeness of graph $G$ is:
\begin{align}
\begin{split}\label{}
C(G) & =  C(C_{n})  \left( 1+ \frac{4-2^{1-p}-2^{1-q}}{n} \right)  
 \\&  \quad    +2p + 2q -3+2^{-p}+2^{-q}+2^{-p-q}.
\end{split}
\end{align}
\end{l2}

\begin{proof}
Let us first connect  cycle  $C_{n}$ with path $P_p$.
The closeness of vertex $p'$, connected to path $P_p$, within cycle  $C_{n}$ is $C(C_{n}) / (n)$.
The closeness of end vertex $p"$ within path  $P_p$ is:
\[
C(p") = 2^{-1} +  2^{-2}+...+ 2^{1-p} = 1-2^{1-p}.
\]
 Using formulae (2) and (4) we receive:
\begin{align}
\begin{split}\label{}
C(C_{n}+ P_p) & = C(C_{n}) +  C(P_p) + 
\left( 1+ \frac{C(C_{n})}{n} \right) \left( 1+ 1-2^{1-p} \right)
\nonumber 
 \\&  =  C(C_{n})  \left( 1+ \frac{2-2^{1-p}}{n} \right) + 2p -2  + 2^{1-p}.
\end{split}
\end{align}
Now we will link graph $C_{n}+ P_p$ to path $P_q$.
Vertex  $q'$, neighbor of vertex $p'$, has closeness (to the vertices of the cycle and to $p$ vertices of the path $P_p$)
within  graph $C_{n}+ P_p$:
\[
C(q')  = \frac{C(C_{n})}{n} + 2^{-2} +  2^{-2}+...+ 2^{-p-1}= \frac{C(C_{n})}{n} + 2^{-1} - 2^{-p-1}.
\]
The closeness of end vertex $q"$ within path  $P_q$ is similar to the closeness of $p"$. Using again formulae (2) and (4) we receive:

\begin{align}
\begin{split}\label{}
C(G) & = C(C_{n}+ P_p) +  C(P_q) 
\nonumber 
 \\&   \quad    \quad  +  \left( 1+ \frac{C(C_{n})}{n} + 2^{-1} - 2^{-p-1} \right)  \left( 1+ 1-2^{1-q} \right).
\\ & = C(C_{n})  \left( 1+ \frac{2-2^{1-p}}{n} \right) + 2p -2  + 2^{1-p}+ 2q -4  + 2^{2-q}
\nonumber 
 \\&   \quad    \quad  +  C(C_{n})   \frac{2-2^{1-q}}{n} 
+ 3 - 2^{-p} - 3 \cdot 2^{-q}
+ 2^{-p-q}
 \\&  =  C(C_{n})  \left( 1+ \frac{4-2^{1-p}-2^{1-q}}{n} \right) +2p + 2q -3
+2^{-p}+2^{-q}+2^{-p-q},
\end{split}
\end{align}
which finishes the proof.
\end{proof}

\section{Minimum of cycles}

Formula (5) could be presented as a function of $p$ and $q$:
\[
C(G) =  C_0 + C_1 (2^{-p}+2^{-q}) +2p + 2q +2^{-p-q}.
\]
If we fix $n$,
in the upper equation $C_0$ is a constant, depending on $n$
 and $C_1$ is a constant:
\[
C_1 = 1 - \frac{2 C(C_{n}) } {n} < 0.
\]
Having $C_1$ less than zero and denoting constant $C_2 = m-n= p + q$,
to find the minimum of $C(G)$ we must solve:
\[
\max \{2^{-p}+2^{-q}\},  \quad  where  \quad p+q = C_2,
\]
or we must find maximum of
\[
F (p)= \{2^{-p}+2^{p-C_2} \}.
\]
Differentiating by $p$ we receive:
\[
F'(p) = -2^{-p-1} lg2 +2^{p-C_2-1} lg2 = 2^{-p-1} lg2 (2^{2p-C_2}-1).
\]
Function $F$ decreases  from $p=0$ to $p=C_2 / 2$ and then increases.
The minimum is when $p=0$ (or $q=0$). From formula (5) and $q=0$ we receive:
\[
\min (C(G)) =  C(C_{n})  \left( 1+ \frac{2-2^{1-p}}{n} \right)  
    +2p + 0 -3+2^{-p}+1+2^{-p}.
\]
\begin{equation}
\label{eq6}
\min (C(G)) =  C(C_{n})  \left( 1+ \frac{2-2^{1-p}}{n} \right)  +2p -2 + 2^{1-p}.
\end{equation}

Let $n=2k$. Then from formula (3) for a cycle with an even number of vertices we receive:
\begin{align}
\begin{split}\label{}
\min(C_e) & = 4k(1-3\cdot 2^{-k-1} ) \left( 1+ \frac{2-2^{1-p}}{2k} \right)  +2p -2 + 2^{1-p}
\nonumber 
\\& = 4k-6k2^{-k}  + (2-3 \cdot 2^{-k} ) (2-2^{1-p}) +2p  -2+ 2^{1-p}
\\& = 4k-6k2^{-k}  +  4 - 2^{2-p} -6 \cdot 2^{-k}  + 3 \cdot 2^{1-k-p}  +2p  -2+ 2^{1-p}
\\& = 4k+2p +  2   -6(k+1)2^{-k}  - 2^{1-p}  + 3 \cdot 2^{1-k-p}. 
\end{split}
\end{align}

\begin{equation}
\label{eq7}
\min (C_e) =  4k+2p +  2   -6(k+1)2^{-k}  - 2^{1-p}  + 3 \cdot 2^{1-k-p}. 
\end{equation}

Differentiating by $k$ we receive:
\begin{align}
\begin{split}\label{}
\min(C_e)'   & = 4  -6\cdot2^{-k}   +6(k+1 )2^{-k-1} lg2   - 3 \cdot 2^{-k-p}lg2
\nonumber 
\\& =  4 +3 \cdot 2^{-k} ( (k+1 ) lg2 -2 - 2^{-p}lg2) > 0.
\end{split}
\end{align}

To have a maximum we need to have a bigger cycle (increase $k$).
The minimum is received when we delete a link from the bigger cycle.
The deleted link can be from both sides of the added link -
hence for the maximum we need both cycles to have (almost) the same length.

Let $n=2k+1$. Then from formulae (6) and (3) for a cycle with an odd number of vertices we receive:
\begin{align}
\begin{split}\label{}
\min(C_o) & = 2(2k+1)(1-2^{-k} ) \left( 1+ \frac{2-2^{1-p}}{2k+1} \right)  +2p -2 + 2^{1-p}
\nonumber 
\\& = 4k+2 -(2k+1)2^{1-k} +  (2-2^{1-k} ) (2-2^{1-p})+2p -2 + 2^{1-p}
\\& = 4k+4 -(2k+3)2^{1-k} -2^{2-p} +  2^{2-k-p}+2p + 2^{1-p}
\\& = 4k+2p+4 -(2k+3)2^{1-k} -2^{1-p} +  2^{2-k-p}. 
\end{split}
\end{align}
\begin{equation}
\label{eq8}
\min (C_o) =  4k+2p+4 -(2k+3)2^{1-k} -2^{1-p} +  2^{2-k-p}. 
\end{equation}

Differentiating by $k$ we receive:
\begin{align}
\begin{split}\label{}
\min(C_o)'   & = 4  - 2^{2-k} +(2k+3)2^{-k} lg2 - 2^{1-k-p}lg2
\nonumber 
\\& =  4 +2^{-k} ((2k+3)lg2 - 4- 2^{1-p}lg2)  > 0.
\end{split}
\end{align}
To have a maximum we again need to have a bigger cycle (increase $k$)
and both cycles to have (almost) the same length.

Now we can prove:
\begin{t5}
The Maximum of minimum closeness of cycle graph is:
\[
Mn (C_m) = \left \{
  \begin{array}{lr}
     8q+2-(2q+3)2^{1-q}  - 2^{2-2q}  + 2^{3-3q},   \quad if \quad m =4q,\\
     8q+6-3(q+2)2^{-q}  - 2^{1-2q}  + 3 \cdot 2^{-3q}, \quad if \quad m =4q+2,\\
     8q+4-(2q+3)2^{1-q}  - 2^{1-2q}  + 2^{2-3q}, \quad if \quad m =4q+1,\\
     8q+8-3(q+2)2^{-q}  - 2^{-2q}  + 3 \cdot 2^{-1-3q}, \quad if \quad m =4q+3.\\
  \end{array}
\right .
\]
\end{t5}
\begin{proof}

Let  $m=4q$. If we connect the  opposite vertices (e.g., $(1,2q+1)$) we
will have a cycle with $2q+1$ vertices and a tail with $2q-1$ vertices (link $(1,4q)$ is deleted). We can use formula (8) with $k=q$ and $p=2q-1$:
\begin{align}
\begin{split}\label{}
C(C_o) & = 4q+2(2q-1)+4 -(2q+3)2^{1-q} -2^{1-2q+1} +  2^{2-q-2q+1}
\nonumber 
\\& =  8q+2-(2q+3)2^{1-q}  - 2^{2-2q}  + 2^{3-3q}. 
\end{split}
\end{align}
If we connect not the opposite vertices (e.g., $(1,2q)$) we
will have a cycle with $2q$ vertices and a tail with $2q$ vertices. We can use formula (7) for cycle with an even number of vertices with $k=q$ and $p=2q$:
\begin{align}
\begin{split}\label{}
C(C_e) & = 4q+4q +  2   -6(q+1)2^{-q}  - 2^{1-2q}  + 3 \cdot 2^{1-q-2q}
\nonumber 
\\& = 8q+  2   -6(q+1)2^{-q}  - 2^{1-2q}  + 3 \cdot 2^{1-3q}. 
\end{split}
\end{align}
Comparing the closenesses with the even and the odd number of vertices we receive:
\begin{align}
\begin{split}\label{}
C_o - C_e & = 8q+2-(2q+3)2^{1-q}  - 2^{2-2q}  + 2^{3-3q}
\nonumber 
\\& \quad -8q- 2  +6(q+1)2^{-q}  + 2^{1-2q}  - 3 \cdot 2^{1-3q}
\\& = 2q2^{-q}-2^{1-2q}+2^{1-3q}=(q-2^{-q})2^{1-q}+2^{1-3q}>0.
\end{split}
\end{align}
This means that the maximum of minimum, when the original graph has $4q$ vertices, is given by connecting opposite vertices $(1,2q+1)$ - the cycle is  with an odd number of vertices.

\quad

Let  $m=4q+2$. If we connect the  opposite vertices (e.g., $(1,2q+2)$) we
will have cycle with $2q+2$ vertices and a tail with $2q$ vertices (link $(1,4q+2)$ is deleted). We can use formula (7)
with $k=q+1$ and $p=2q$:
\begin{align}
\begin{split}\label{}
C(C_e) & = 4(q+1)+4q +  2   -6(q+1+1)2^{-q-1}  - 2^{1-2q}  + 3 \cdot 2^{1-q-1-2q}
\nonumber 
\\& = 8q+6-3(q+2)2^{-q}  - 2^{1-2q}  + 3 \cdot 2^{-3q}. 
\end{split}
\end{align}
If we connect not the opposite vertices (e.g., $(1,2q+1)$) we
will have a cycle with $2q+1$ vertices and a tail with $2q+1$ vertices. We can use formula (8) for cycles with an odd number of vertices with $k=q$ and $p=2q+1$:
\begin{align}
\begin{split}\label{}
C(C_o) & = 4q+2(2q+1)+4 -(2q+3)2^{1-q} -2^{1-2q-1} +  2^{2-q-2q-1}
\nonumber 
\\& = 8q+6-(2q+3)2^{1-q}  - 2^{-2q}  + 2^{1-3q}. 
\end{split}
\end{align}
Comparing the closenesses with the even and the odd number of vertices we receive:
\begin{align}
\begin{split}\label{}
C_e - C_o & = 8q+6-3(q+2)2^{-q}  - 2^{1-2q}  + 3 \cdot 2^{-3q}
\nonumber 
\\& \quad -8q-6+(2q+3)2^{1-q} + 2^{-2q}  - 2^{1-3q}
\\& = q2^{-q}- 2^{-2q}+2^{-3q}=(q-2^{-q})2^{-2q}+2^{-3q}>0.
\end{split}
\end{align}
This means that the maximum of minimum, when the original graph has $4q+2$ vertices, is given by connecting the opposite vertices $(1,2q+2)$ - the cycle is  with an even number of vertices.

\quad 

The cases with the original graph having odd
number of vertices is considered  similar way.
Let  $m=4q+1$.  If we connect one pair of  close to opposite vertices (e.g., $(1,2q+1)$) we will have cycle with $2q+1$ vertices and a tail with $2q$ vertices (link $(1,4q+1)$ is deleted). We can use formula (8) with $k=q$ and $p=2q$:
\begin{align}
\begin{split}\label{}
C(C_o) & = 4q+4q+4 -(2q+3)2^{1-q} -2^{1-2q} +  2^{2-q-2q}.
\nonumber 
\\& =  8q+4-(2q+3)2^{1-q}  - 2^{1-2q}  + 2^{2-3q}. 
\end{split}
\end{align}
If we have the other pair of  close to opposite vertices (e.g., $(1,2q+2)$) the 
deleted link could be $(1,2)$ and the result will be the same.
This means that the maximum of minimum, when the original graph has $4q+1$ vertices, is given by connecting vertices $(1,2q+1)$ - the cycle is  with odd number of vertices.

\quad 

Let  $m=4q+3$.  If we connect one pair of  close to opposite vertices (e.g., $(1,2q+2)$) we will have cycle with $2q+2$ vertices and a tail with $2q+1$ vertices (link $(1,4q+3)$ is deleted). We can use formula (7) with $k=q+1$ and $p=2q+1$:
\begin{align}
\begin{split}\label{}
C(C_o) & = 4(q+1)+2(2q+1) +  2   -6(q+1+1)2^{-q-1} 
\nonumber 
\\&  \quad  - 2^{1-2q-1}  + 3 \cdot 2^{1-q-1-2q-1}
\\& =  8q+8-3(q+2)2^{-q}  - 2^{-2q}  + 3 \cdot 2^{-1-3q}. 
\end{split}
\end{align}

If we have the other pair of  close to opposite vertices (e.g., $(1,2q+3)$) the 
deleted link could be $(1,2)$ and the graph is symmetric  to the upper one.
This means that the maximum of minimum, when the original graph has $4q+3$ vertices, is given by connecting vertices $(1,2q+2)$ - the cycle is  with an even number of vertices.
\end{proof}

To find a formula for other criteria for cycles is more complicated.

\section{Conclusion}
We have considered decision theory for closeness of utility networks.
If for a given period  one link will be destroyed  (by disaster) we try
to find the best link to be built, so the damages are minimized.
We use the closeness of the network as a payment function and 
apply the decision theory to determine our best action. 

We use different graphs (cycle, lollipop graphs, 2 connected
complete graphs) and different criteria to determine which  
link to build.

Future research could be focused  on calculating all other criteria for different graphs,
including cycle graphs, lollipop graphs, etc.

\quad

\noindent  \textbf {Data Availability}

\noindent All data are incorporated into the article.

\end{document}